\documentclass[letterpaper,twocolumn,english,aps,prb,superscriptaddress]{revtex4-1}
\usepackage[T1]{fontenc}
\usepackage[latin9]{inputenc}
\usepackage{amsmath}
\usepackage{amssymb}
\usepackage{graphicx}

\makeatletter

\pdfpageheight\paperheight
\pdfpagewidth\paperwidth

\usepackage{sidecap}

\makeatother

\usepackage{babel}
\begin{document}
\title{Correlated nonequilibrium steady states without energy flux}
\author{Hristiana Atanasova}
\affiliation{Institut für Theoretische Physik, Universität Hamburg, Jungiusstraße
9, D-20355 Hamburg, Germany}
\affiliation{School of Chemistry, Tel Aviv University, Tel Aviv 6997801, Israel}
\author{Alexander I. Lichtenstein}
\affiliation{Institut für Theoretische Physik, Universität Hamburg, Jungiusstraße
9, D-20355 Hamburg, Germany}
\author{Guy Cohen}
\email{gcohen@tau.ac.il}

\affiliation{School of Chemistry, Tel Aviv University, Tel Aviv 6997801, Israel}
\affiliation{The Raymond and Beverley Sackler Center for Computational Molecular
and Materials Science, Tel Aviv University, Tel Aviv 6997801, Israel}
\date{\today}
\begin{abstract}
Floquet engineering of closed quantum systems can lead to the formation
of long-lived prethermal states that, in general, eventually thermalize
to infinite temperature. Coupling these driven systems to dissipative
baths can stabilize such states, establishing a true nonequilibrium
steady state. We demonstrate that in a certain strongly interacting
lattice model coupled to a bath and driven by an electric field, such
steady states can have the remarkable property that the cycle-averaged
rate of energy transfer between the lattice and the baths vanishes.
Despite this, we show that these states retain a clear nonequilibrium
nature.

\end{abstract}
\maketitle

\section{Introduction}

Closed, interacting quantum systems driven by a time-dependent external
field are generically expected to absorb energy from the field and
evolve towards an infinite temperature state.\citep{rigol_thermalization_2008}
However, in certain cases they can be trapped in a nonequilibrium
prethermal state \citep{bukov_prethermal_2015,bukov_schrieffer-wolff_2016,herrmann_floquet_2017,weidinger_floquet_2017}
that is long-lived, but not permanent.\citep{dalessio_long-time_2014}
In contrast, in cases ranging from, \emph{e.g.,} a single particle
in a double well \citep{grossmann_coherent_1991} or lattice \citep{dunlap_dynamic_1986}
to discrete few-state models \citep{grosmann_localization_1992} and
systems of noninteracting particles,\citep{cadez_dynamical_2017}
periodic fields can lead to complete dynamical localization or coherent
destruction of coupling such that special Floquet states can survive
indefinitely.

Many-body localized systems \citep{gornyi_interacting_2005,basko_metalinsulator_2006}
are perhaps the only known example of extended, isolated and interacting
systems that do not absorb heat when driven by a periodic potential
.\citep{ponte_periodically_2015} The signatures of many-body localized
phases can even survive the introduction of phonons when driven.\citep{lenarcic_activating_2018,lenarcic_critical_2019}
However, localization is known to be fragile \citep{dalessio_long-time_2014}
and is rapidly destroyed in the presence of generic small perturbations.\citep{luitz_absence_2017}
Recent experimental progress in, \textit{e.g.} ultrafast optical manipulation
of solids \citep{forst_nonlinear_2011} and control of ultracold atomic
gases \citep{gross_quantum_2017} has brought these questions, and
periodically driven many-body systems in general, to the forefront
of present-day scientific discussion.\citep{moessner_equilibration_2017,bordia_periodically_2017,reitter_interaction_2017,pomarico_enhanced_2017,messer_floquet_2018,roy_anomalous_2018,werner_entropy-cooled_2019}

Real systems are never perfectly isolated. Open quantum systems can
dissipate absorbed energy into baths, and therefore typically reach
a nonequilibrium steady state (NESS) depending on the properties of
both the coupling to the bath and the drive.\citep{weiss_quantum_1999}
A dissipative bath allows for stabilization of nonthermal Floquet
states in nonintegrable many-body lattice models.\citep{tsuji_nonequilibrium_2009,murakami_nonequilibrium_2017,murakami_nonequilibrium_2018,peronaci_resonant_2018}
At NESS, heating still occurs, but the rate at which energy is absorbed
from the field is equal to the rate at which it is dissipated to the
bath.

With a few recent exceptions,\citep{murakami_nonequilibrium_2017,murakami_nonequilibrium_2018,qin_nonequilibrium_2018}
the rate of energy dissipation into the bath from many-body lattice
systems driven into periodic NESS has been largely unexplored. We
present a case where this rate, averaged over time, appears to be
either zero or negligibly small. Notably, maintaining the nonequilibrium
state in the model considered here does require a constant absorption
of energy, which must then be dissipated. However, even though the
lattice is driven, energy is absorbed and dissipated \textit{entirely
by the baths} (which are also driven for reasons explained below).
Within any single field cycle, the amount of energy flowing out of
the lattice is exactly balanced by the amount flowing into it.
\begin{figure}
\begin{centering}
\includegraphics[width=8.6cm]{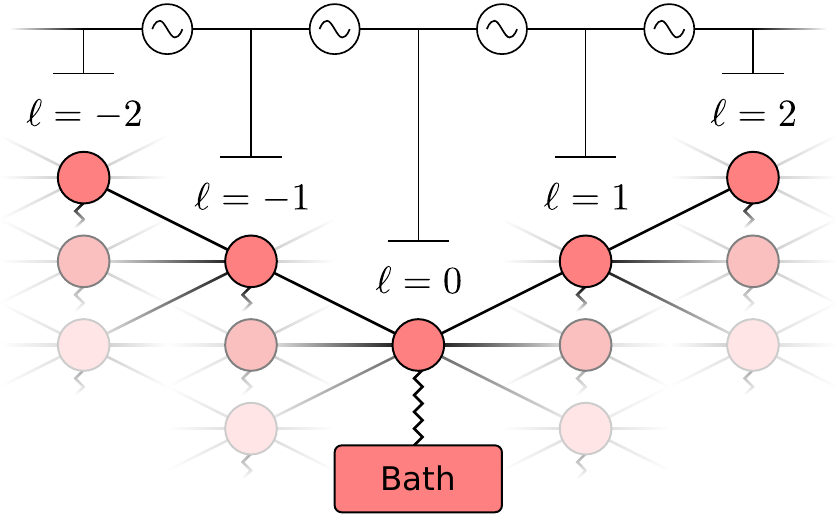}
\par\end{centering}
\caption{Model: sites on a Bethe lattice are divided into layers, and coupled
to the sites in adjacent layers and to baths. A time-dependent potential
difference $\Delta\phi\left(t\right)$ is applied between consecutive
layers, driving a current within the lattice.\label{fig:model}}
\end{figure}

\section{Model}

We consider a system for which exact dynamics can be obtained by numerical
methods, though the present treatment will be approximate: a Hubbard
model on a Bethe lattice of coordination number $Z$, at the infinite
$Z$ limit, driven by an oscillating electric field and in contact
with a set of noninteracting metallic baths (see Fig.~\ref{fig:model}).
Each bath is coupled to a single lattice site and held at the same
potential as that site (bath sites are therefore necessarily driven).
The system's Hamiltonian is
\begin{equation}
\hat{H}=\hat{H}_{L}+\hat{H}_{B}+\hat{H}_{LB}.\label{eq:hamiltonian}
\end{equation}
Here, the particle\textendash hole symmetric lattice Hamiltonian $\hat{H}_{L}$
and bath Hamiltonian $\hat{H}_{B}$ are given by
\begin{equation}
\begin{aligned}\hat{H}_{L}\left(t\right) & =-J\sum_{\left\langle ij\right\rangle \sigma}\hat{d}_{i\sigma}^{\dagger}\hat{d}_{j\sigma}+\sum_{i,\sigma}\phi_{i}\left(t\right)\hat{n}_{i\sigma}\\
 & +U\sum_{i}\left(\hat{n}_{i\uparrow}-\frac{1}{2}\right)\left(\hat{n}_{i\downarrow}-\frac{1}{2}\right),\\
\hat{H}_{B} & =\sum_{ik\sigma}\left(\varepsilon_{k}+\phi_{i}\left(t\right)\right)\hat{b}_{ik\sigma}^{\dagger}\hat{b}_{ik\sigma}.
\end{aligned}
\label{eq:lattice_and_bath_H}
\end{equation}
The $\hat{d}_{i\sigma}$ and $\hat{d}_{i\sigma}^{\dagger}$, respectively,
are fermionic annihilation and creation operators on the lattice and
$\hat{n}_{i\sigma}=\hat{d}_{i\sigma}^{\dagger}\hat{d}_{i\sigma}$.
The $\hat{b}_{ik\sigma}$ and $\hat{b}_{ik\sigma}^{\dagger}$ are
a corresponding set of fermionic operators within the noninteracting
bath attached to lattice site $i$. $J\equiv\frac{t_{0}}{\sqrt{Z}}$
is the lattice hopping amplitude $t_{0}$ scaled by the coordination
number and $U$ is the Hubbard interaction strength. We set $t_{0}\equiv1$
and employ it as our unit of energy, also setting $\hbar\equiv1$.
The system is subjected to an oscillating electric potential $\phi_{i}\left(t\right)$,
and the $\varepsilon_{k}$ are single-particle energies within the
baths.

The final term in the Hamiltonian describes the lattice\textendash bath
coupling, and is given by
\begin{equation}
\hat{H}_{LB}=\sum_{ik\sigma}\left(V_{k}\hat{b}_{ik\sigma}^{\dagger}\hat{d}_{i\sigma}+h.c.\right).
\end{equation}
The $V_{k}$ are chosen for numerical convenience so as to generate
a flat density of states $\Gamma_{i}\left(\omega\right)=\sum_{k}\left|V_{k}^{2}\right|\delta\left(\omega-\varepsilon_{k}\right)=\frac{\Gamma}{\left(1+e^{\nu\left(\omega-\omega_{c}\right)}\right)\left(1+e^{-\nu\left(\omega+\omega_{c}\right)}\right)}$
with bandwidth $2\omega_{c}$ and edges softly falling off over an
energy range $\frac{1}{\nu}$. We take $\omega_{c}=50$, $\nu=10$
(essentially the wide band limit) and $\Gamma=\frac{1}{2}$. We further
set the baths' inverse temperature to $\beta=1$ and their chemical
potential $\mu$ to zero. We note that modeling the baths by Lindblad
operators at this point would completely fail to capture the physics
in which we are interested, because we will consider energy going
\emph{back and forth} between the lattice and baths; a non-Markovian
description is therefore crucial.

To set up a nonequilibrium situation, the system is divided into layers
labeled by an index $\ell$ such that each site in layer $\ell$ is
connected to $\frac{Z}{2}$ neighbors in each of the layers $\ell\pm1$,
and $\phi_{i}=\ell_{i}E\cos\left(\varOmega t\right)$. We note that
each lattice site is always at the same potential as its bath, as
implied by the requirement that the bath describes a spatially adjacent
metallic region. The potential difference between two consecutive
layers is therefore $\Delta\phi\left(t\right)=\pm E\cos\left(\varOmega t\right)$,
and drives an alternating current in the lattice. Fig.~\ref{fig:model}
illustrates this at $Z=6$; at $Z=2$ the system would simply be a
1D chain in a linear oscillating field. While $\phi_{i}\left(t\right)$
breaks translation symmetry, we follow other studies \citep{aoki_nonequilibrium_2014}
in employing a Peierls substitution to eliminate the potential terms
from the Hamiltonian, shifting them to a site-independent phase in
the hopping amplitudes. This is equivalent to a rotating frame transformation
with respect to $\hat{R}\left(t\right)=e^{-i\sum_{i\sigma}\Phi_{i\sigma}\left(t\right)\left(\hat{n}_{i\sigma}+\sum_{k}\hat{b}_{ik\sigma}^{\dagger}\hat{b}_{ik\sigma}\right)}$,
where $\Phi_{i\sigma}\left(t\right)=\int_{0}^{t}\phi_{i\sigma}\left(\tau\right)\mathrm{d\tau}=l_{i}A\sin(\varOmega t)$
and $A\equiv\frac{E}{\Omega}$ is a dimensionless driving amplitude.

\section{Numerical solution and observables}

We evaluate the time evolution of the system within nonequilibrium
dynamical mean field theory (DMFT), which is \textit{exact} for models
of this type at $Z\rightarrow\infty$.\citep{georges_dynamical_1996,freericks_nonequilibrium_2006,aoki_nonequilibrium_2014}
DMFT maps the lattice model onto an effective impurity model, which
we solve within strong coupling perturbation theory using the noncrossing
approximation (NCA).\citep{bickers_review_1987,eckstein_nonequilibrium_2010,cohen_greens_2014,cohen_greens_2014-1,antipov_currents_2017}
This approach is appropriate in the insulating regime, where the Coulomb
interaction is strong compared to the lattice hopping, and is the
\emph{only} approximation used here. The lattice is initially assumed
to be in an antiferromagnetic Néel state, and is coupled to the bath
at time $t=0$. We evaluate the double occupation $d_{i}(t)=\left\langle \hat{n}_{i\uparrow}\left(t\right)\hat{n}_{i\downarrow}\left(t\right)\right\rangle $;
the local Green's functions $G_{\sigma}\left(t,t^{\prime}\right)=\left\langle \hat{d}_{i\sigma}\left(t\right)\hat{d}_{i\sigma}^{\dagger}\left(t^{\prime}\right)\right\rangle $;
and the lattice\textendash bath energy current
\begin{equation}
\begin{aligned}I_{E}(t) & =-\frac{\mathrm{d}}{\mathrm{d}t}\left\langle \hat{H}_{B}\left(t\right)-\hat{H}_{B,\mathrm{internal}}\left(t\right)\right\rangle \\
 & =-i\sum_{ik\sigma}\varepsilon_{k}V_{k}\left\langle \hat{b}_{ik\sigma}^{\dagger}\hat{d}_{i\sigma}-\hat{d}_{i\sigma}^{\dagger}\hat{b}_{ik\sigma}\right\rangle .
\end{aligned}
\end{equation}
The latter definition is given by the time derivative of the bath
energy after removing the internal energy generation $\frac{\mathrm{d}}{\mathrm{d}t}\left\langle \hat{H}_{B,\mathrm{internal}}\left(t\right)\right\rangle =\sum_{i,k,\sigma}\dot{\phi_{i}}\left(t\right)\hat{b}_{ik\sigma}^{\dagger}\hat{b}_{ik\sigma}$.
This term describes irreversibly dissipated energy, which will not
be explored here.

We note that the model of Eq.~\eqref{eq:hamiltonian} was used for
theoretical convenience rather than experimental realizability. One
could generalize the theoretical treatment to more realistic finite-dimensional
lattices, at the cost that DMFT will no longer be exact. It would
then further be possible, at the cost of greatly increased computational
difficulty, to improve the theory by introducing nonlocal corrections
to DMFT in any of a variety of established methods.\citep{maier_quantum_2005,rubtsov_dual_2008,kananenka_systematically_2015,zgid_finite_2017}
This goes beyond the scope of the present manuscript.

\section{High frequency limit}

Before considering numerical results, a qualitative understanding
of the system can be obtained from analytical considerations. Bukov
\textit{et al.}~\onlinecite{bukov_schrieffer-wolff_2016} showed
that that for periodic driving of period $T$ or equivalently frequency
$\varOmega=\frac{2\pi}{T}$, such that $\hat{H}\left(t\right)=\hat{H}\left(T+t\right)$,
an approximate static Hamiltonian $\hat{H}_{eff}$ can be derived
at the high frequency limit (HFL) by way of a generalized Schrieffer\textendash Wolff
transformation and an expansion in $\frac{1}{\Omega}$. In the present
model, the transformation can be performed by going to the rotating
frame with respect to the operator $\hat{R}\left(t\right)=e^{-iUt\sum_{i}\left(\hat{n}_{i\uparrow}-\frac{1}{2}\right)\left(\hat{n}_{i\downarrow}-\frac{1}{2}\right)-i\sum_{i\sigma}\Phi_{i\sigma}\left(t\right)\left(\hat{n}_{i\sigma}+\sum_{k}\hat{b}_{ik\sigma}^{\dagger}\hat{b}_{ik\sigma}\right)}$.
For resonant driving $U=l\varOmega$ with $l\in\mathcal{N}$, the
leading-order effective Hamiltonian $\hat{H}_{\mathrm{eff}}^{\left(0\right)}=\frac{1}{T}\int_{0}^{T}\mathrm{d}t\hat{H}_{\mathrm{rot}}\left(t\right)$
then takes the simple form

\begin{equation}
\begin{aligned}\hat{H}_{\mathrm{eff}}^{\left(0\right)} & =-J\sum_{\left\langle ij\right\rangle \sigma}\left[\mathcal{J}_{0}\hat{g}_{ij\sigma}+\left(\mathcal{J}_{l}\left(-1\right)^{l\eta_{ij}}\hat{h}_{ij\sigma}^{\dagger}+h.c.\right)\right]\\
 & -\frac{4i}{\pi}\sum_{ik\sigma}V_{k}\left[\hat{h}_{ik\sigma}^{\dagger}-h.c.\right].
\end{aligned}
\label{eq:Heff}
\end{equation}
Here, $\eta_{ij}=1$ for $i>j$ and $\eta_{ij}=0$ for $i<j$. $\mathcal{J}_{l}\equiv\mathcal{J}_{l}\left(A\right)$
denotes the $l^{\mathrm{th}}$ order Bessel function of the dimensionless
driving amplitude $A=\frac{E}{\varOmega}$. The effective Hamiltonian
within the lattice contains two types of operators with couplings
set by $A$: the first is doublon and holon hopping, described by
$\hat{g}_{ij\sigma}=\hat{n}_{i\bar{\sigma}}\hat{d}_{i\sigma}^{\dagger}\hat{d}_{j\sigma}\hat{n}_{j\bar{\sigma}}+\left(1-\hat{n}_{i\bar{\sigma}}\right)\hat{d}_{i\sigma}^{\dagger}\hat{d}_{j\sigma}\left(1-\hat{n}_{j\bar{\sigma}}\right)$;
and the second is creation/annihilation of doublon\textendash holon
pairs, described respectively by $\hat{h}_{ij\sigma}^{\dagger}=\hat{n}_{i\bar{\sigma}}\hat{d}_{i\sigma}^{\dagger}\hat{d}_{j\sigma}\left(1-\hat{n}_{j\bar{\sigma}}\right)$
and its Hermitian conjugate $\hat{h}_{ij\sigma}$. Coupling to the
bath introduces modified terms of the second type, given by $\hat{h}_{ik\sigma}^{\dagger}=\hat{b}_{ik\sigma}^{\dagger}\hat{d}_{i\sigma}\left(\hat{n}_{i\bar{\sigma}}-\frac{1}{2}\right)$
and $\hat{h}_{ik\sigma}$ and respectively denoting creation / annihilation
of excitations associated with lattice\textendash bath tunneling processes.
The energy current operator, in the rotating frame and up to first
order in the high frequency expansion, is $\hat{I}_{E,\mathrm{eff}}^{(0)}\left(t\right)=\frac{4}{\pi}\sum_{i,k,\sigma}\varepsilon_{k}V_{k}\left[\hat{h}_{ik\sigma}^{\dagger}+h.c.\right]$,
and contains only doublon/holon density dependent hopping terms.

We make no particular claims regarding the accuracy of the HFL (corrections
in the resonant case have been explored in the literature \citep{herrmann_floquet_2017}),
and our main results do not rely on taking this limit. Nevertheless,
we show below that it captures the basic physical picture at a qualitative
(but not quantitative) level.

We proceed by writing a master equation for the NESS occupation probabilities
$P_{\alpha}$ of a lattice site. The index $\alpha$ refers to the
four possible states of an isolated site, denoted by $\left|0\right\rangle $
(unoccupied), $\left|\sigma\right\rangle $ (singly occupied with
$\sigma\in\left\{ \uparrow,\downarrow\right\} $) and $\left|\uparrow\downarrow\right\rangle $
(doubly occupied). The static NESS must have no net flow in or out
of any state, such that $\sum_{\beta}R_{\alpha\rightarrow\beta}P_{\alpha}=\sum_{\beta}R_{\beta\rightarrow\alpha}P_{\beta}.$
Given the rates $R_{\alpha\rightarrow\beta}$, the $P_{\alpha}$ can
be obtained from this condition with the additional requirement of
normalization, $\sum_{\alpha}P_{\alpha}=1$.

We estimate rates in the HFL by projecting Eq.~\ref{eq:Heff} onto
the subspace of a pair of adjacent lattice sites $i$ and $j$, $\left|n_{i}n_{j}\right\rangle $;
and considering Golden rule transition rates between the different
states. The expectation values of particle number operators $\hat{n}_{j\mathbf{\sigma}}$
or their products are then expressed in terms of local state probabilities,
\textit{i.e.} $\left\langle \hat{n}_{j\sigma}\right\rangle =P_{\left|\sigma\right\rangle _{j}}+P_{\left|\uparrow\downarrow\right\rangle _{j}}$.
For example, the rate $R_{\left|\sigma\right\rangle \leftarrow\left|0\right\rangle }=-\mathcal{J}_{0}\left(A\right)P_{\left|\sigma\right\rangle }P_{\left|0\right\rangle }-\mathcal{J}_{l}\left(A\right)P_{\left|\uparrow\downarrow\right\rangle }P_{\left|0\right\rangle }$.
To this, one must add the rates $R_{\left|0\right\rangle \leftarrow\left|\sigma\right\rangle }^{\mathrm{bath}}=R_{\left|\uparrow\downarrow\right\rangle \leftarrow\left|\sigma\right\rangle }^{\mathrm{bath}}=\frac{2\Gamma}{\pi^{2}}f$
and $R_{\left|\sigma\right\rangle \leftarrow\left|0\right\rangle }^{\mathrm{bath}}=R_{\left|\sigma\right\rangle \leftarrow\left|\uparrow\downarrow\right\rangle }^{\mathrm{bath}}=\frac{2\Gamma}{\pi^{2}}\left(1-f\right)$
for transitions mediated by tunneling of electrons between the lattice
and bath, where $f=\frac{1}{1+e^{\beta\left(\varepsilon-\mu\right)}}$
is the bath Fermi function.

The rate equations can be further simplified by enforcing the symmetries
$P_{\left|0\right\rangle }=P_{\left|\uparrow\downarrow\right\rangle }$
and $P_{\left|\sigma\right\rangle }=P_{\left|\bar{\sigma}\right\rangle }$,
but remain cubic, and their solution is unwieldy. To gain more intuition,
we linearize the equations by replacing the probabilities \textit{within
the rates} by their approximate values at large $U$: $P_{\left|\uparrow\downarrow\right\rangle }^{\mathrm{rate}}\simeq0$
and $P_{\left|\sigma\right\rangle }^{\mathrm{rate}}\simeq\frac{1}{2}$.
This finally yields the double occupation probability $d^{\mathrm{HFL}}$:

\begin{align}
d^{\mathrm{HFL}}\equiv P_{\left|\uparrow\downarrow\right\rangle } & =\frac{\pi^{2}\left|\mathcal{J}_{l}\left(A\right)\right|+4\left(1-f\right)}{8\left(\frac{\pi^{2}\left|\mathcal{J}_{l}\left(A\right)\right|}{4}+1\right)},\label{eq:hfl_me_probabilities}
\end{align}
where $f$ is once again the bath Fermi function parameterized by
$\beta=1$, $\mu=0$ and the transition energy $\varepsilon=5$. Within
the steady state master equation, state $\alpha$ is occupied at probability
$P_{\alpha}$ and a transition from it to state $\beta$, which occurs
at rate $R_{\beta\leftarrow\alpha}^{\mathrm{bath}}$, contributes
the energy difference $E_{\beta}-E_{\alpha}$ to the the energy flux.
The lattice\textendash bath energy current is therefore $I_{E}^{\mathrm{HFL}}=-\sum_{\beta}\left(E_{\beta}-E_{\alpha}\right)R_{\beta\leftarrow\alpha}^{\mathrm{bath}}P_{\alpha}$,
with $E_{\alpha}$ the energy of the isolated state $\left|\alpha\right\rangle $.
An analogous discussion for deriving the charge current can be found
in Ref.~\onlinecite{datta_quantum_2005}.

\section{Results}

\begin{figure}
\includegraphics{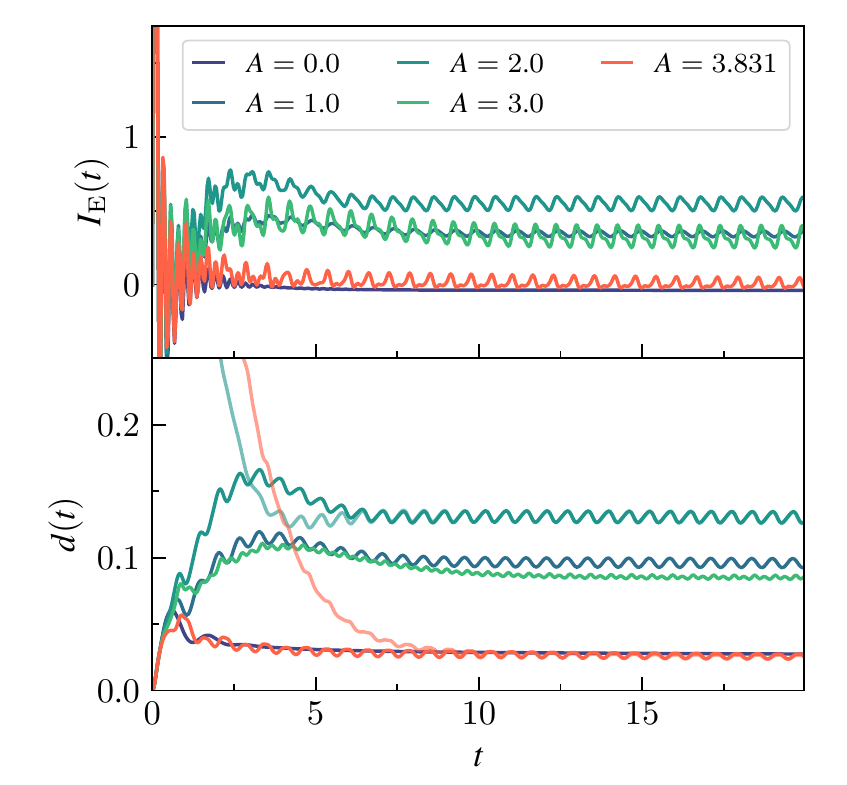}

\caption{Energy current between the lattice and bath (upper panel) and double
occupation (lower panel) within the NCA. The electric field's frequency
$\varOmega=U=10$ is set to resonance, and several field amplitudes
$A$ are shown. Lighter lines denote an alternative initial condition
(see text).\label{fig:Time-dependent-energy-current}}
\end{figure}

In Fig.~\ref{fig:Time-dependent-energy-current}, we plot the NCA
dot\textendash bath energy current $I_{E}(t)$ (upper panel) and double
occupation $d\left(t\right)$ (lower panel), which show the numerical
data extrapolated to $\Delta t\rightarrow0$. Both quantities are
shown at resonant driving $l=1$ (\textit{i.e.} $\varOmega=U=10$)
and at several field amplitudes $A$. In equilibrium ($A=0$), energy
flows between the system and bath only for times $t\lesssim2$, while
the system is still relaxing to equilibrium with the bath after the
initial quench. Activating the electric field induces oscillations
in both observables. A stable NESS emerges at all amplitudes after
a timescale $t\gtrsim2\sim\frac{1}{\Gamma}$, and we have verified
that doubling the final simulation time makes essentially no difference
(not shown). At small amplitudes ($A=1,2$), $I_{E}(t)$ and $d\left(t\right)$
both grow with increasing $A$; however, at higher amplitudes ($A\gtrsim3$)
they decrease with increasing $A$, with a very strong suppression
apparent at the first Bessel zero $A\approx3.831$. Since the system
is close to the Mott regime at $A=0$, an increase in $d(t)$ implies
that energy from the field is absorbed by doublons and holons.

Floquet stabilization differs from dynamical decoupling in that the
NESS is determined by the driving and dissipative coupling rather
than the initial condition. To demonstrate this, lighter lines in
Fig.~\ref{fig:Time-dependent-energy-current} show, for two representative
amplitudes, dynamics starting from a state where even(odd) lattice
layers are doubly occupied(unoccupied). The doublon/holon occupation
rapidly attains the same NESS to within numerical accuracy.

\begin{figure}
\includegraphics{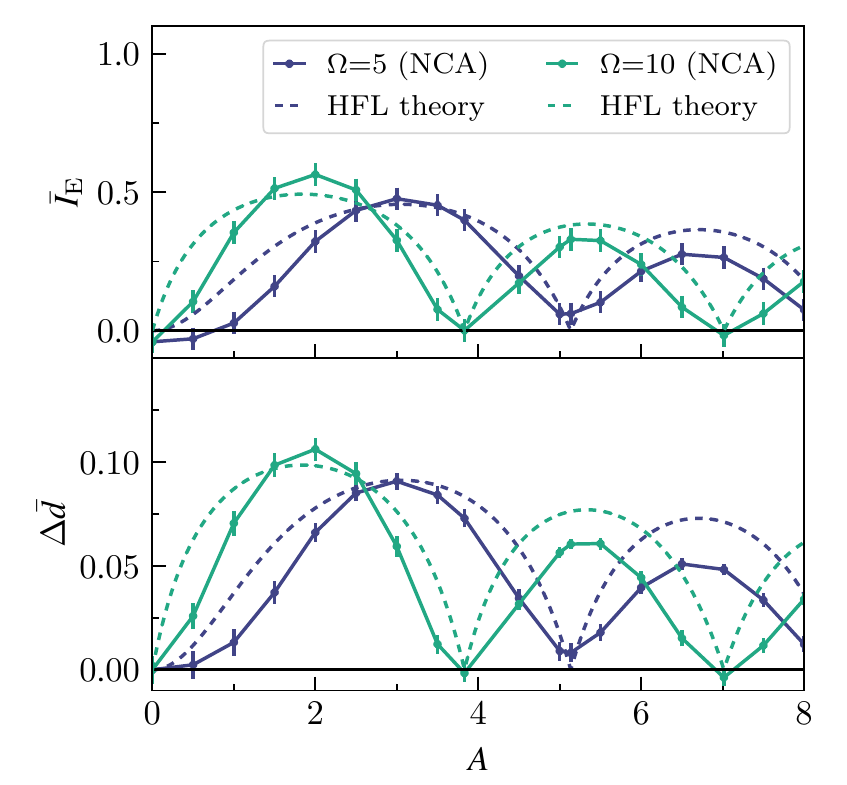}

\caption{Amplitude dependence of the period averaged energy current $\bar{I}_{E}(A)$
(upper panel) and double occupation difference $\Delta\bar{d}(A)$
(lower panel, see text for definition). The driving frequencies shown
are $\varOmega=U/2=5$ and $\varOmega=U=10$. NCA results are plotted
as solid lines, while dashed lines denote values obtained from the
HFL theory and multiplied by a factor of 50 to be visible on the same
scale.\label{fig:period-averaged-IE-and-d}}
\end{figure}

A simpler picture emerges if we consider NESS observables averaged
over a driving cycle. \textit{\emph{For any observable $O$, we define}}
$\bar{O}\equiv\frac{1}{T}\int_{t_{m}-T}^{t_{m}}\mathrm{d}t\thinspace O\left(t\right)$,
where $t_{m}$ is the maximum simulation time. The upper panel of
Fig.~\ref{fig:period-averaged-IE-and-d} then shows the NCA period
averaged energy current $\bar{I}_{E}(A)$, at $\Omega=U$ ($l=1$)
and at $\Omega=\frac{U}{2}$ $\left(l=2\right)$, as solid lines.
The error bars denote numerical and finite-time uncertainties in our
solution of the NCA equations. These are obtained by considering deviations
from exact sum rules: for the energy current the error estimate is
given by the difference of $\bar{I}_{E}\left(A=0\right)$ from zero,
and for the double occupation by the deviation of the trace of the
reduced density matrix in the atomic subspace from one. The lower
panel shows $\varDelta\bar{d}(A)=\bar{d}(A)-\bar{d}(0)$, the difference
between the period averaged double occupation $\bar{d}(A)$ and its
equilibrium counterpart at $A=0$. Both $\bar{I}_{E}(A)$ and $\varDelta\bar{d}(A)$
have an amplitude dependence reminiscent of a Bessel function $\left|\mathcal{J}_{l}\left(A\right)\right|$,
with $l$ set by the frequency. The similarity between $\bar{I}_{E}$
and $\bar{d}$ suggests that dissipation is mediated by the creation
and destruction of holons and doublons. Remarkably, in the $l=1$
regime, tuning the dimensionless amplitude $A$ to Bessel zeros suppresses
the creation of doublon\textendash holon pairs and therefore the average
rate of energy transfer between the lattice and the baths vanishes
to within the numerical uncertainties. Fig.~\ref{fig:period-averaged-IE-and-d}
demonstrates that for amplitudes with $A\approx3.831,7.015$ a NESS
is maintained such that within a cycle and within numerical uncertainties,
no energy dissipation occurs. At $l=2$, finite dissipation is predicted,
but it should be noted that the NCA is less reliable in this regime.

These results can be understood qualitatively within the analytical
HFL theory, shown as dotted lines in Fig.~\ref{fig:period-averaged-IE-and-d}.
While this simple theory captures the suppression effect well, its
prediction differs from that of the NCA both in the shape of the curves
and by a quantitative factor of $\sim50$ in the overall size of observables.
Furthermore, our HFL theory predicts full suppression also at $l=2$,
whereas the NCA does not.

\begin{figure}
\includegraphics{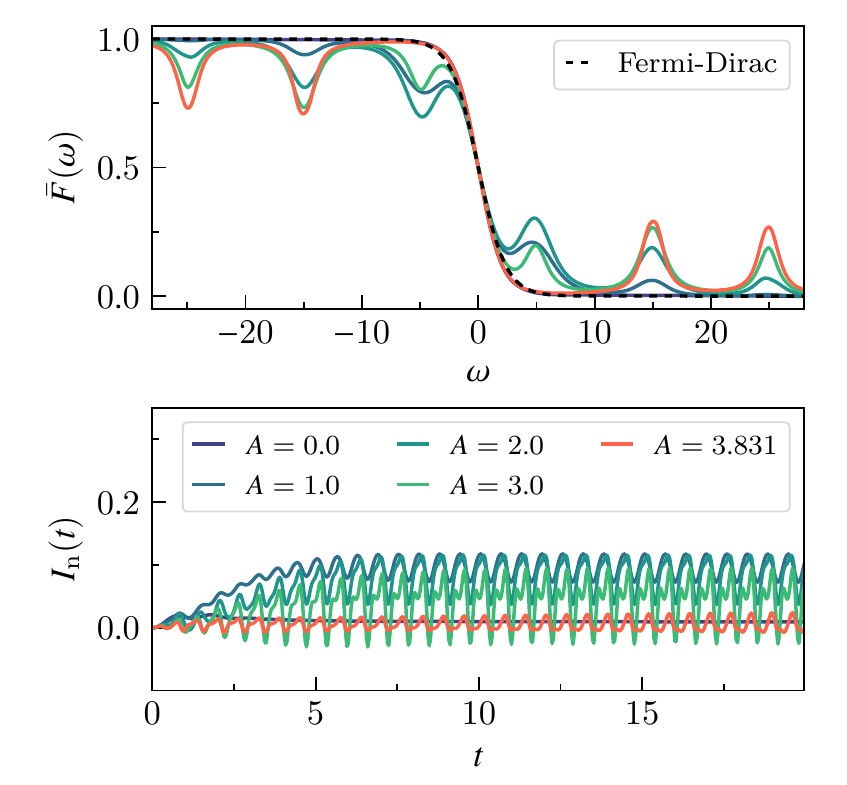}\caption{Upper panel: nonequilibrium distribution functions for a resonant
electric field with $\varOmega=U=10$ and a series of amplitudes $A$.
The dotted line is the Fermi\textendash Dirac distribution at the
bath temperature $\beta=1$. Lower panel: particle current in the
lattice. For $A=3.831$ a net flow within the lattice and a nonequilibrium
distribution remain clearly visible, such that the system cannot be
in an effective equilibrium state.\label{fig:distribution-function-and-particle-current}}
\end{figure}

The counterintuitive nature of a dissipation-free NESS might lead
one to conclude that the system may be at an effective equilibrium
state when driven at Bessel zero amplitudes. Interestingly, this is
not the case. To show this, we plot the period averaged nonequilibrium
distribution function $\bar{F}\left(\omega\right)=-\frac{i}{2}\frac{\bar{G}^{<}\left(\omega\right)}{\Im\bar{G}^{R}\left(\omega\right)}$
at $l=1$ in the top panel of Fig.~\ref{fig:distribution-function-and-particle-current}.
In equilibrium $\bar{F}(\omega)$ matches the Fermi\textendash Dirac
distribution $f\left(\omega\right)$ at the bath temperature. Driving
the system modulates $\bar{F}(\omega)$ such that resonances / antiresonances
appear at $\omega=-\frac{U}{2}\pm n\varOmega$ with $n\in\mathcal{N}$,
corresponding to $n$-photon absorption processes. At an amplitude
matching the first Bessel zero single-photon absorption is suppressed,
but the spectrum retains a clear nonequilibrium shape. The lower panel
of Fig.~\ref{fig:distribution-function-and-particle-current} shows
the particle current $I_{n}(t)$ in the lattice at the same parameters.
In equilibrium $I_{n}(t)$ vanishes within numerical uncertainty,
as it must; but this happens at no other amplitude, including the
Bessel zero case. We note that the time-averaged lattice current can
be nonzero, but cancels with the lattice\textendash bath current in
agreement with particle conservation.

\section{Conclusion}

We considered an infinite-dimensional interacting many-body lattice
model driven by periodic fields and coupled to a bath. For this model,
we showed that it is possible to engineer a family of driven steady
states with distinct nonequilibrium characteristics, but without time-averaged
dissipation of energy between the system and bath. At the limit of
high frequency driving we showed that this surprising result can be
understood at the level of a simple analytical theory. Our numerical
results are based on dynamical mean field theory, exact for this model;
and on the noncrossing approximation at a physical regime where it
is thought to be reliable. A variety of interesting questions can
now be raised: since the noncrossing approximation can be lifted by
a variety of modern numerical methods,\citep{cohen_taming_2015,antipov_currents_2017,dong_quantum_2017,profumo_quantum_2015,bertrand_quantum_2019,bertrand_reconstructing_2019,moutenet_cancellation_2019}
it will be of some interest to see whether this result survives an
exact treatment. Another important challenge is to check whether the
effect remains present for more realistic finite-dimensional models
that may also feature nonlocal interactions, vibrational degrees of
freedom and disorder. Finally, theoretical work will be needed to
understand how generic this result is, and whether it can it be used
to control dissipation and selectively engineer desired steady states
in practical applications.
\begin{acknowledgments}
G.C. acknowledges support by the Israel Science Foundation (Grant
No. 1604/16), the German Academic Exchange Service (DAAD) and the
Minerva Foundation. The authors would like to thank Martin Eckstein
and Yevgeny Bar-Lev for illuminating and useful scientific discussions.
\end{acknowledgments}

\bibliographystyle{apsrev4-1}
\bibliography{Library}

\end{document}